# An integrated experimental and computational framework for modeling creep behavior in shale rocks induced by chemo-mechanical loading

Ravi Prakash[1], Sara Abedi[*,1,2]

[1]Department of Civil Engineering, Texas A&M University, College Station, TX 77843, United States

[2]Department of Petroleum Engineering, Texas A&M University, College Station, TX 77843, United States

*sara.abedi@tamu.edu

**Abstract**

Creep deformation in shale rocks is an important factor in many applications, such as the sustainability of geostructures, wellbore stability, evaluation of land subsidence, $CO_2$ storage, toxic waste containment, and hydraulic fracturing. One mechanism leading to this time-dependent deformation under a constant load is the dissolution/formation processes accompanied by chemo-mechanical interactions with a reactive environment. When dissolution/formation processes occur within the material phases, the distribution of stress and strain within the material microstructure changes. In the case of the dissolution process, the stress carried by the dissolving phase is distributed into neighboring voxels, which leads to further deformation of the material. The aim of this study was to explore the relationship between the microstructural evolution and time-dependent creep behavior of rocks subjected to chemo-mechanical loading. This work uses the experimentally characterized microstructural and mechanical evolution of a shale rock induced by interactions with a reactive brine ($CO_2$-rich brine) and a non-reactive brine ($N_2$-rich brine) under high-pressure and high-temperature conditions to compute the resulting time-dependent deformation using a time-stepping finite-element-based modeling approach. Sample microstructure snapshots were obtained using segmented micro-CT images of the rock samples before and after the reactions. Coupled nanoindentation/EDS provided spatial alteration of the mechanical properties of individual material phases due to the dissolution and precipitation processes as a result of chemo-mechanical loading of the samples. The time-dependent mechanically informed microstructures were then incorporated into a mechanical model to calculate the creep behavior caused by the dissolution/precipitation processes independent of the inherent viscous properties of the mineral phases. The results indicate the substantial role of the dissolution/precipitation processes on the viscous behavior of rocks subjected to reactive environments.

# 1. INTRODUCTION

Shale rocks exhibit significant creep deformation over spatial and temporal scales, which has important implications for various applications, including the sustainability of geostructures, wellbore stability (Horsrud et al., 1994), evaluation of land subsidence (Abel & Lee, 1980; Chang & Zoback, 2008), toxic waste containment, and hydraulic fracturing (Fjaer et al., 2008; Rybacki et al., 2017). Several experimental and computational studies suggest that the dissolution of load-bearing phases occurring during the evolution of the microstructure plays a major role in causing viscoelastic/viscoplastic (VE/VP) deformation in multiphasic composites, such as geomaterials (Buscarnera & Das, 2016; Hu & Hueckel, 2007a, 2007b; Hueckel et al., 2016; Karato & Jung, 2003; Rutter, 1976; Schimmel et al., 2022). This microstructural evolution could be the result of chemical reactions that occur as a result of interactions with a reactive environment, causing stress/strain redistribution (Ghoussoub & Leroy, 2001; Karato & Jung, 2003; Niemeijer et al., 2002; Rutter, 1976, 1983; Shimizu, 1995; Weyl, 1959).

Various theoretical models and computational algorithms have been developed to study the role of chemical-mechanical couplings on the stress/strain distribution in porous materials. Continuum thermodynamic models have been used extensively to study the long-term mechanical alteration induced by reactive fluid flow in rocks, for instance, by exploring the dominant mechanisms involved in dissolution/precipitation processes or by studying the solid-fluid phase interfacial transformation due to chemical reactions (Buscarnera & Das, 2016; Ghoussoub & Leroy, 2001; Heidug & Leroy, 1994; Lehner, 1990; Tengattini et al., 2014). Models developed within the context of Poromechanics have been used to study the chemo-mechanics of porous materials, resulting in notable progress in resolving the chemo-mechanical properties of cementitious materials, such as chemically induced mechanical damage and early age mechanical growth (Coussy, 2004; Ulm & Coussy, 1995, 1996; Ulm et al., 1999). Progress has also been made in the coupling of chemical reactions and the mechanical properties of geomaterials based on the adjustment of plasticity theory to chemical loading (Gajo et al., 2015; Gajo et al., 2002; Hu & Hueckel, 2007a; Hueckel et al., 2016). In this context, chemical variables are embodied in the plastic hardening rules such that reactions give rise to expansion/contraction of the elastic domain. This approach has been effective in describing different types of mechanisms, including time-dependent deformation induced by changes in pore fluid chemistry, coupled damage, and dissolution-induced creep of geomaterials (Castellanza et al., 2008; Nova et al., 2003).

From an experimental standpoint, several studies have been performed on the time-dependent mechanical behavior of rocks owing to chemo-mechanical effects (Baud et al., 2009; Clark & Vanorio, 2016; Dautriat et al., 2011; Grgic, 2011; Neveux et al., 2014). Static and dynamic mechanical measurements were performed and geomechanical models were proposed (Castellanza & Nova, 2004; Ciantia et al., 2015; Fernandez-Merodo et al., 2007; Vanorio, 2015). Studies have shown that the extent of rock mechanical alteration and the resulting creep deformation due to chemo-mechanical interactions are influenced by the rock composition, fluid chemistry, microstructural and transport properties of the rock, temperature, pressure, reaction time, and specific surface area (Cook et al., 2011; Haldar, 2013; Nermoen et al., 2016; Rimstidt et al., 2017). The chemical reactions that occur between rocks and reactive fluids cause microstructural alterations in the rock frame, which lead to short- and long-term debonding processes (Ciantia & Castellanza, 2016), alteration of stress patterns, and formation of microcracks (Vanorio, 2018). Combined with mechanical loading, these chemo-mechanical interactions can cause additional deformation owing to the pressure solution and compaction (Atkinson, 1984; Dewers & Ortoleva, 1990; He et al., 2002; Tada & Siever, 1989; Wawersik et al., 2001).

These studies have played a vital role in leading investigations and designing a wealth of predictive models (Vanorio, 2015). However, some past studies are subject to limitations owing to the macroscopic nature of their investigation. Even when considering pore-scale effects, most conventional models and experimental studies consider the effect of chemo-mechanical loading to be spatially uniform over the bulk scale of the rock. These limitations motivated this study, in which we captured the true microstructural evolution of the rock as a result of interactions with reactive fluids and obtained time-dependent mechanical and deformational behavior at the microscale.

This study made use of experimentally characterized microstructural and mechanical evolution of shale rock induced by interaction with a reactive brine ($CO_2$-rich brine) and a non-reactive brine ($N_2$-rich brine) under high-pressure and high-temperature conditions to compute the resulting time-dependent deformation using a time-stepping finite-element-based modeling approach. Digital snapshots generated by micro-computed tomography (micro-CT) imaging before and after the reactions provided a microstructural model of the shale sample. Coupled nanoindentation and energy dispersive X-ray spectroscopy (EDS) revealed the spatial alteration of the mechanical properties of different material phases caused by dissolution and precipitation processes due to the chemo-mechanical loading of the rock samples. The evolution of the VE/VP properties of the samples was then estimated by combining the microstructural evolution with the stress/deformation distribution within the material with respect to time. For this purpose, the kinematic framework developed by Li et al. (2016) and Kannan and Rajagopal (2011) was adopted to link the microstructural evolution to the change in the stress and strain fields, and dissolution/precipitation-induced creep as a result of rock-fluid interaction. By carrying out calculations on samples reacted with reactive and non-reactive brines, the effects of chemical reactions and the resulting microstructural changes on the time-dependent deformation of the studied samples were investigated.

## 2. MATERIAL AND METHODS

### 2.1. Materials

Creep deformation was calculated for Permian shale samples to illustrate the implementation and applicability of the adopted model. Rock-eval pyrolysis of the Permian sample indicated an organic content of 5.15%. The mass percent composition according to X-ray diffraction analysis showed that the major constituents of the sample were quartz (54.59%), illite (23.31%), and oligoclase feldspar (16.61%). The sample also consisted of dolomite (2.54%), pyrite (1.67%), and siderite (1.27%).

## 2.2. Experimental Procedure

The rock samples were exposed to $CO_2$-rich ($CO_2$ condition) and $N_2$-rich brine ($N_2$ condition) in a titanium-made Parr batch dissolution reactor. The samples were submerged in 1 molar NaCl synthetic brine at 100°C and 12.5 MPa for 2 weeks. The $N_2$ condition represented the control condition without chemical attacks of the acidic brine and was achieved by injecting $N_2$ into the reactor. Before and after the reaction experiment, high-resolution images of the sample volume with a voxel size of 3 µm were obtained to characterize the microstructural evolution of the rock. Variations in the X-ray attenuation coefficient were used to segment the various constitutive minerals. The Avizo software was used for image processing and segmentation. Image stack preprocessing and background removal were performed using the steps described by Prakash et al. (2021). Moreover, coupled nanoindentation and EDS were performed on cross-sectional samples before and after the reaction to track the evolution of the mechanical properties of individual material phases as a function of depth from the exposed surface (Prakash et al., 2021). Fig. 1 shows the results obtained from micro-CT imaging in terms of the volume fraction of different mineral phases as a function of the distance from the exposed surface. Fig. 2 shows the variation in the indentation modulus after 14 days of reaction for the clay-rich and quartz-rich phases along the depth of reaction obtained from the coupled nanoindentation and EDS techniques. Each point in Fig. 2 is the average indentation modulus of a particular mineral phase (clay- or quartz-rich phases) within each column of the nanoindentation grid. The micro-CT results (Fig. 1) indicate the dissolution of feldspar and precipitation of clay close to the reacted surface for samples reacted under $CO_2$ condition. The quartz particles were also dissolved close to the reacted surface, accompanied by quartz precipitation owing to feldspar transformation in an acidic environment in the sample that reacted under $CO_2$ condition. Under $N_2$ condition, Al-rich zones were observed close to the reacted surface. These Al-rich zones were combined with clay-rich zones in the micro-CT image analysis. Such Al-rich zones are observed under neutral pH conditions and result from preferential leaching of Si from feldspar, leaving Al-rich zones in feldspars (Huang & Keller, 1970; Huang & Kiang, 1972; Kawano & Tomita, 1996; Prakash et al., 2021; Reesman & Keller, 1965, 1968).

Both the clay-rich and quartz-rich phases are weakened and become more porous by the dissolution of mineral grains for the entire depth of analysis (710 µm) under $CO_2$ condition. This dissolution of mineral grains is accompanied by the secondary mineral precipitation of clay and quartz closer to the reacted surface, resulting in a higher modulus of these phases in a narrow zone adjacent to the exposed surface. These precipitations result from the dissolution of feldspar and precipitation of kaolinite and phyllosilicate minerals such as smectite, illite (Hangx & Spiers, 2009; Lu et al., 2013), and quartz grains as the reaction proceeds (Yuan et al., 2019).

For the sample that reacted under $N_2$, Fig. 2b shows a high-intensity reaction close to the reacted surface, limited to a depth of approximately 200 µm. This decreased modulus can be attributed to weathering and washing of grains close to the surface, making the sample more porous and softer. After a depth of 200 μm, the indentation moduli of both the quart-rich and clay-rich phases matched those of the unreacted sample. The reader is referred to (Prakash et al., 2023) for more details on the experimental procedure.

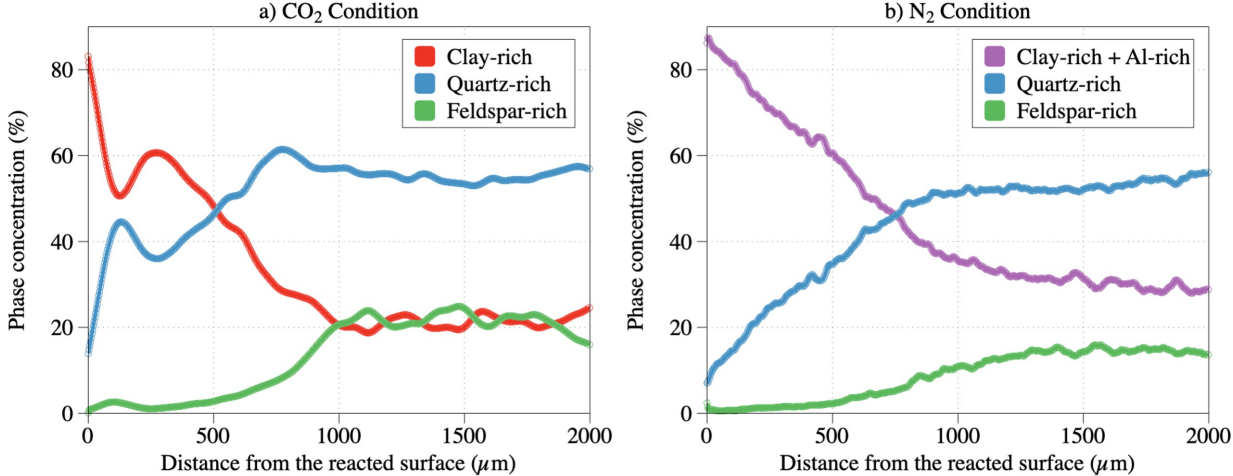

Figure 1- Variation of volume fraction of different mineral phases as a function of distance from the reacted surface for samples reacted under a) $CO_2$ condition and b) $N_2$ condition.

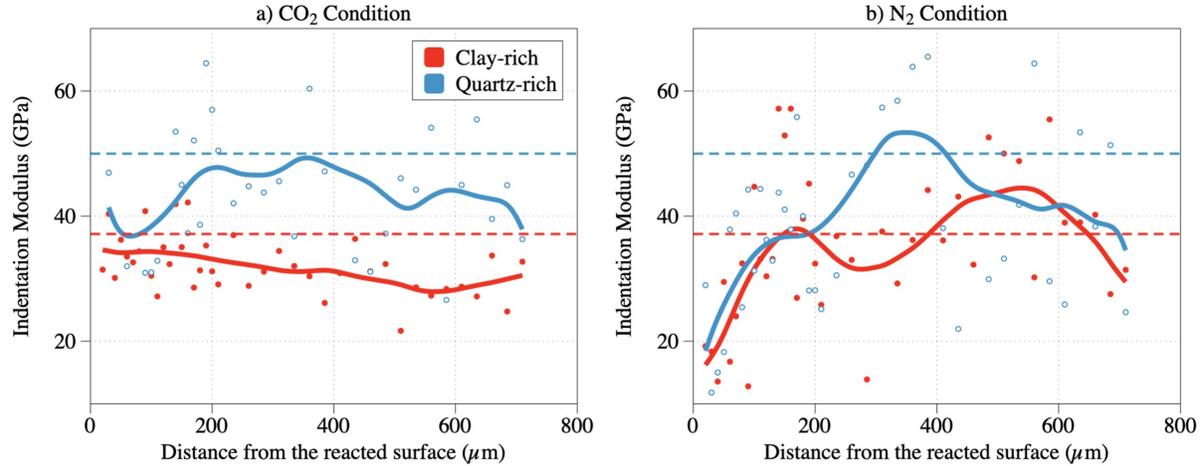

Figure 2- Variation of indentation modulus as a function of distance from the reacetd surface for clay-rich and quartz-rich phases for samples reacted under a) $CO_2$ condition and b) $N_2$ condition. Each point shows the average indeintation modulus of clay-rich or quartz-rich phases whitin each column of indentation grid. The solid colored lines represent the overal trend. The dottted lines show the average of indentation moudlus of the unreacted phase.

## 2.3. Computational Methodology

This study integrated an experimentally informed microstructure model and a time-stepping finite-element mechanical model. The microstructure model produces evolving sample microstructures based on the results of micro-CT analysis combined with coupled nanoindentation and SEM/EDS, which are then used in the model to predict the evolution of the deformation and elastic properties of samples over time. More details regarding the utilized methodology are provided in the following sections.

### 2.3.1. Microstructure Model

The microstructural evolution from unreacted to reacted samples for both $CO_2$ and $N_2$ samples was achieved by (1) updating the mineral composition according to the micro-CT analysis and (2) updating the mechanical properties according to the coupled indentation and SEM/EDS data (Prakash & Abedi, 2022). As mentioned in Sec. 2.2, micro-CT imaging was performed on unreacted and reacted samples, and then segmentation analysis was performed on those images. To reduce the computational cost and for simplicity, a 2D microstructure with a size of 2400 μm × 1800 μm (800 × 600 pixels) was selected for FEM analysis. The 2D segmented image was selected such that it was representative of the complete volume and had the same composition as the complete volume.

The model used in the analysis obtained its geometry directly from the microstructure, and the size of the elements in the finite element method was governed by the size of the voxels in the micro-CT images (3 μm). The geometry was discretized such that each pixel was an element. The term pixel, instead of voxel, is used here because it is a 2D analysis. Each pixel (consisting of a unique phase) was a four-node bilinear square element. It should be noted that increasing image resolution or the number of voxels per simulation box length can enhance the accuracy of model predictions (Haecker et al., 2005). In this study, we employed the smallest element size, which corresponds to a voxel. As noted earlier, the 2D segmented image was selected such that it was representative of the complete volume and shared the same composition. Reducing element size below a voxel would refine the mesh but exponentially increase the number of elements, necessitating the selection of a smaller micro-CT image due to high computational costs. However, as explained in Sec. 2.3.1, we refrain from selecting a smaller image since it would not accurately represent the complete volume of the sample. Fig. 3 shows the unreacted microstructure of the sample, which contains various mineral phases. This microstructure was a cropped image from one of the segmented image slices of the unreacted sample.

As mentioned in Sec. 2.2, the variation in the concentration of the various mineral phases in relation to the distance to the exposed surface for the reacted samples was obtained for the entire 3D segmented volume (Fig. 1). To generate the microstructure of the reacted samples, microstructural changes were made to the unreacted microstructure. Using the unreacted microstructure, binary images were extracted for each mineral, with black pixels showing that mineral and white regions showing the rest of the microstructure. Fig. 4 shows the binary image for the feldspar-rich phases. This binary image was used for the erosion operation. Because the volume proportion of feldspar reduces due to dissolution, eroding operations were applied to feldspar-rich phases. The volume fraction of feldspar-rich phases changed with the depth of reaction; hence, different amounts of erosion are required at different locations. To achieve this, the unreacted microstructure was divided into 12 zones (Fig. 5). The size of these zones was decided based on the intensity of the reaction and the extent of microstructural alteration.

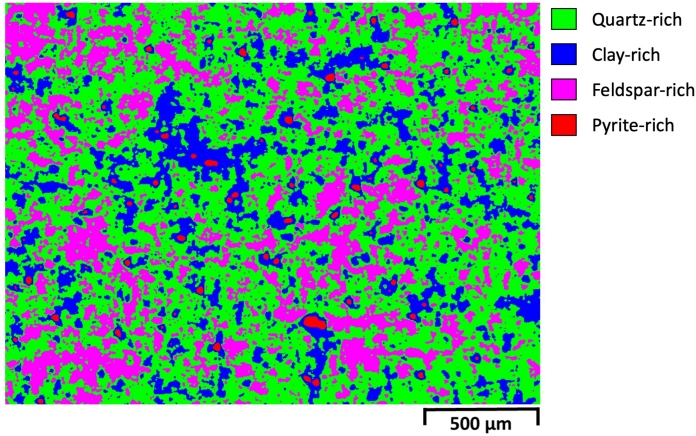

Figure 3- Microstructure of unreacted Permian shale showing different mineral phases. 2D segmented image is selected such that it is representative of the complete volume.

Each of the first four zones was 50 × 1800 μm in size, with a longer edge parallel to the exposed surface. The next seven zones were 100 μm × 1800 μm in size. No changes were made to the microstructure of the last (twelfth) zone, representing the unreacted zone. The first four zones were intentionally selected as thinner than the next seven zones because of the sharp variations in the volume fractions closer to the reacted surface, which can be effectively captured using thinner zones. The smallest zone thickness was restricted to 50 μm to maintain continuity in the microstructure, and the maximum thickness was limited to 100 μm to capture the changes in the volume fraction of different material phases (i.e., feldspar-, quartz-, and clay-rich phases) in the microstructure. For each zone of the unreacted microstructure, the average volume percentage of feldspar was calculated, and the erosion operation was applied such that the volume percentage decreased to the average value for the reacted sample in that region. Multiple iterations of the erosion operation were performed to achieve this value. The freeware image analysis software ImageJ was used to perform erosion/dilation operations (Ferreira & Rasb, 2012). Once the erosion operation was performed for all the zones individually, all images were stitched together to obtain a full-size image for the feldspar-rich phases.

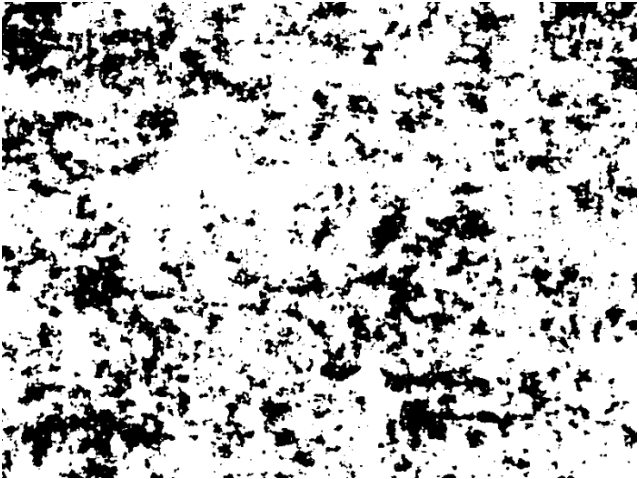
Figure 4- Binary image of feldspar-rich phases obtained from the unreacted microstructure.

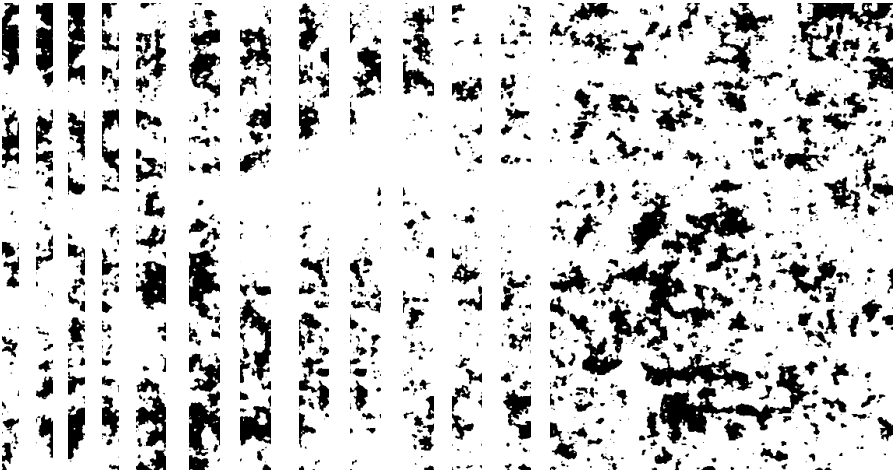
Figure 5- Binary image of unreacted feldspar-rich phases divided into 12 zones for erosion/dilation operations.

Eroding operation on the grains was performed for quartz-rich phases close to the reacted surface (dissolution zone), whereas the grains were dilated in certain regions next to the dissolution zone, where the volume fraction of quartz-rich phases slightly increased according to micro-CT results (Fig. 1a). The erosion/dilation procedure and the number/thickness of zones were kept the same as for the feldspar-rich regions to obtain a full-size image for quartz mineral. No changes were made to the pyrite-rich phases; hence, the reacted binary image for pyrite-rich phases was the same as the unreacted binary image of this mineral. Once successfully obtained, all full-size binary images of reacted feldspar-rich, quartz-rich, and pyrite-rich minerals were superimposed together. The remaining regions on the superimposed images were marked as clay. Therefore, the dissolution of feldspar and quartz (cumulative dissolution) resulted in the dilation of the clay regions and increased the volume fraction of clay in the microstructure. Fig. 6 shows the resultant microstructure of the Permian sample after reacting with $CO_2$ condition and $N_2$ condition for 14 days.

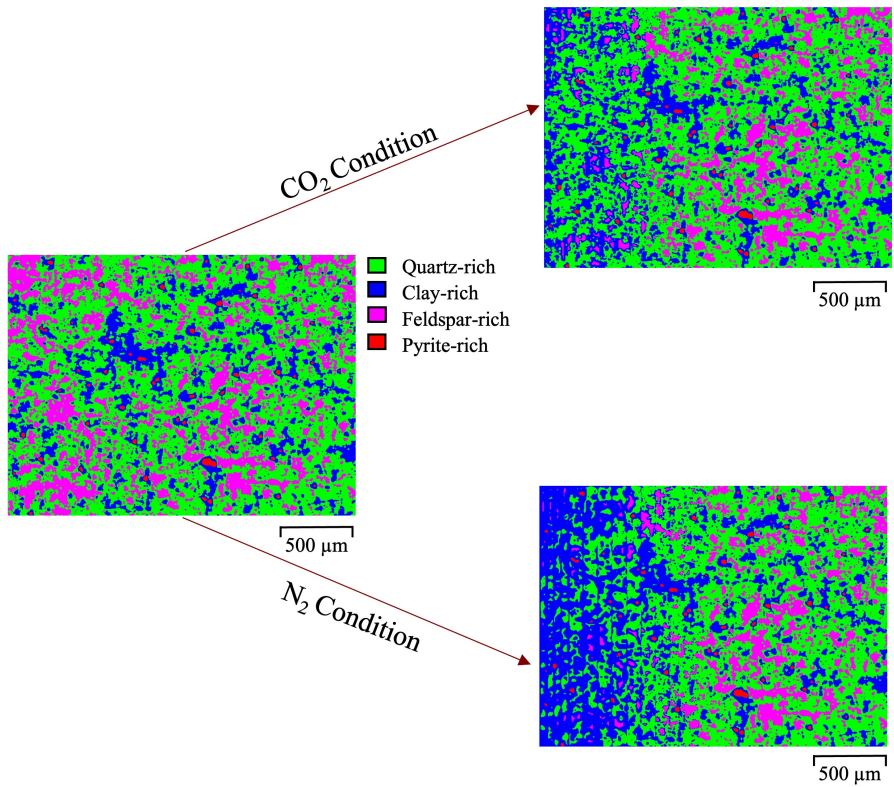

Figure 6- Microstructure of unreacted Permian shale along with reacted microstructures after 14 days of reaction under $CO_2$ and $N_2$ conditions, showing different mineral phases, produced using image analysis. The reacted surface is on the left side of the images. For the sample reacted under $N_2$ condition, clay-rich zones include Al-rich zones as well.

The effect of the growth of reaction depth over time on creep deformation was also examined for the sample reacted under $CO_2$ condition. For this purpose, after obtaining the microstructure of the 14-day reaction sample, the microstructure was generated at larger time steps with the assumption that the reaction depth is a function of time, and it follows Fick's law relationship (Fick, 1855), that is,

$$D(t_i) = k\sqrt{t_i} \qquad (1)$$

where $D(t_i)$ is the reaction depth at any time $t_i$, and $k$ is a constant of proportionality. The thickness of each zone was calculated by assuming the above relationship. For example, the zones with a thickness of 50 μm for 14 days increased to a thickness of 70.71 μm after 28 days of reaction. Similarly, the zones with a thickness of 100 μm increased to 141.42 μm. Notably, this part of the study only considers the effect of reaction depth growth; therefore, no changes were made to the volume fraction of minerals within each reaction zone after any time $t_i$.

Based on these two assumptions, microstructures were generated after 28, 42, and 56 days of the reaction. The steps of the zoning and erosion/dilation methodology remained the same as for the 14-day microstructure. The only change was in the thickness of each zone, depending on the reaction time. Fig. 7 shows the microstructure under unreacted condition and after 14, 28, 42, and 56 days of reaction.

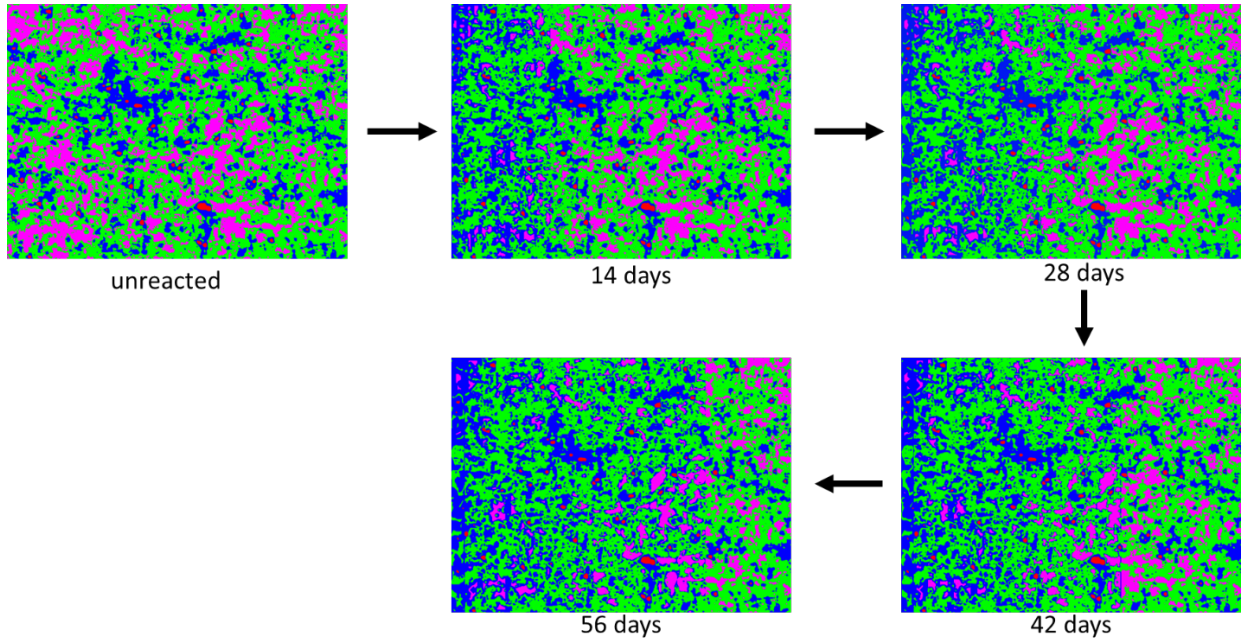

Figure 7- Microstructure of unreacted Permian shale and after 14, 28, 42 and 56 days of progression of the reaction depth under $CO_2$ condition, generated using image analysis procedures. The reacted surface is on the left side of the images.

### 2.3.2. Assigning Mechanical Properties

As mentioned previously, coupled nanoindentation and EDS performed on the polished cross-section of the reacted samples were used to obtain the mechanical properties of the individual phases as a function of the reaction depth (Fig. 2). The obtained indentation modulus was assigned to the microstructure. The average indentation moduli of the clay- and quartz-rich phases were calculated for each zone, and the calculated moduli were assigned to the minerals in the specific zones. The indentation modulus of feldspar and pyrite for the reacted microstructure was assumed to be the same as that of the unreacted sample because of the lack of sufficient indentation points on the feldspar- and pyrite-rich phases in the reacted sample. This is a good assumption attributed to the small compositional volumes of these minerals. This assumption simplified the microstructure. This change in the modulus from unreacted to reacted samples was assumed to be constant after 28, 42, and 56 days of reaction. Notably, the thickness of each zone increased at each time step; thus, although the indentation modulus of each depth remained constant, its variation changed with time.

### 2.3.3. Conceptualization of the Adopted Kinematic Model

When an elastic material subjected to the dissolution process was exposed to external loading, the entire composite and every voxel/element of the material were deformed from the original configuration to the current configuration. Once the load was removed, the composite did not return to its original configuration, but changed to its natural configuration. The evolution of the natural configuration has been found to cause creep in composite bodies (Kannan & Rajagopal, 2011; Li et al., 2016; Rajagopal & Srinivasa, 2004).

The complete evolution of the configuration that occurs with the composite also occurs with individual voxels of the composite. The natural configuration of a composite may vary from the natural configuration of the voxel. This difference in motion is referred to as the apparent strain in the composite (Li et al., 2016). Fig. 8 provides a graphical depiction of this concept (Li et al., 2016; Rajagopal & Srinivasa, 2004). In the context of continuum mechanics, apparent strain is defined as the strain that is present regardless of any external stimuli. Previous studies have demonstrated that the presence of apparent strain is a significant factor contributing to the irreversible dissolution/precipitation-induced creep. In order to accurately predict the composite material's overall deformation behavior, it is necessary to capture the impact of apparent strain on the inherent stress and strain fields within the composite (Li et al., 2016; Wineman & Rajagopal, 2000).

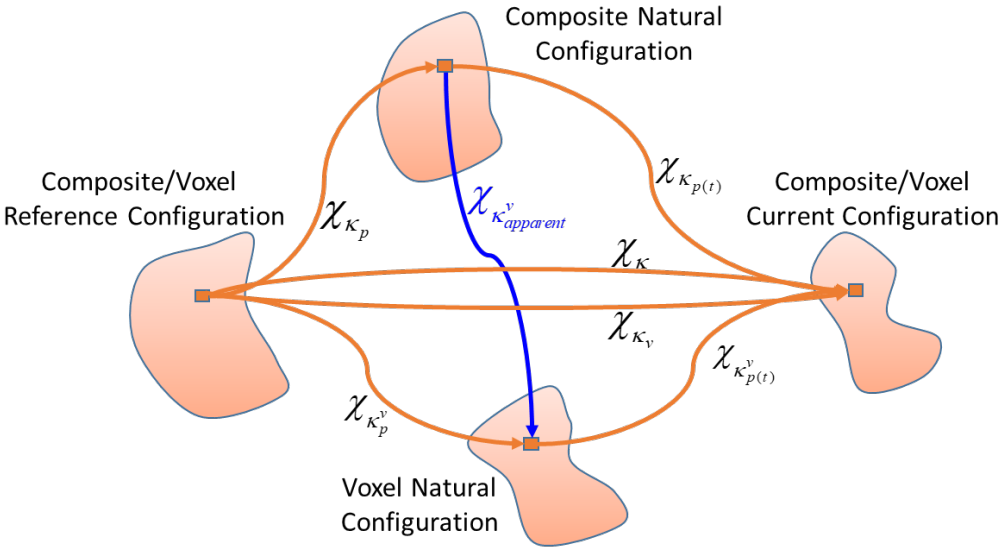

Figure 8- A schematic view showing the computational framework (adopted form (Li et al., 2016)).

Creep deformation in the multiphasic composite was computed by quantifying the spatially averaged strain field and the spatially averaged stress field within the composite with certain boundary conditions. For a composite body, the spatially averaged stress and strain field can be computed based on the three deformation gradients $F_{k_{p(t)}}$, $F_{k_{apparent}^v}$, and $F_{K_{p(t)}^v}$ related to the motions $\chi_{k_{p(t)}}$, $\chi_{k_{apparent}^v}$, and $\chi_{K_{p(t)}^v}$, respectively. By means of these deformation gradients, the Green-Lagrangian strain tensor generated for the motion $\chi_{k_{p(t)}}$ can be expressed as (Li et al., 2016)

$$E_{k_{p(t)}} = \frac{1}{2}\left(\left(F_{k_{apparent}^v}F_{k_{p(t)}^v}\right)^T \left(F_{k_{apparent}^v}F_{k_{p(t)}^v}\right) - I\right) \quad (2)$$

where $I$ denotes the identity tensor. Eq. (2) can also be written as a function of displacement as:

$$E_{k_{p(t)}} = \left(\text{Grad}_{k_p^v}u_{k_p^v}^T + I\right)E_{k_{apparent}^v}\left(\text{Grad}_{k_p^v}u_{k_p^v} + I\right) + E_{k_{p(t)}^v} \quad (3)$$

where $\text{Grad}_{k_p^v}$ represents the gradient operation with respect to the natural voxel configuration. Considering infinitesimal and linearized strains $\varepsilon$, Eq. (3) can be written as

$$\varepsilon_{K_{p(t)}} = \varepsilon_{K_{apparent}^v} + \varepsilon_{K_{p(t)}^v} \text{ and } \varepsilon = \varepsilon_{apparent} + \varepsilon_{mechancial} \quad (4)$$

where $\varepsilon_{apparent}$ corresponds to the apparent strain induced by the dissolution-deformation process (i.e., due to chemical reactions), indicating that the $\varepsilon_{apparent}$ encompasses the strain brought about by the changes in rock microstructure linked to chemical reactions. $\varepsilon_{mechancial}$ is the mechanical strain of the voxel resulting from the stress to which the voxel has been subjected.

### 2.3.4. Finite Element Calculation

A finite element-based homogenization model, elas2d.f, was used to predict the elastic moduli of the composites from 2D reconstructed digital images (Garboczi, 1998). This computational scheme uses different microstructural configurations of the composite at different time steps as inputs and includes the stress distribution effects of the dissolution process. The concept of time steps was included in this computational scheme to account for the historical dependence of the mechanical properties of the material. For a given microstructure at a given time step, the total mechanical energy stored in the entire microstructure should be minimized to achieve mechanical equilibrium. For this analysis, each voxel/element was assumed to behave as a purely linearly elastic material. Then, at each time step, the total elastic energy stored in one voxel/element is given by the following equation when the composite is subjected to infinitesimal strain:

$$E_n = \frac{1}{2}\iiint_0^1 \varepsilon_{pq}C_{pqrs}\varepsilon_{rs}\,dxdydz \quad (5)$$

where $E_n$ is the total stored elastic energy, $\varepsilon_{pq}$ and $C_{pqrs}$ are the infinitesimal strain tensor and the elastic moduli in full tensorial form, $p, q, r, s = 1, 2,$ or $3$, respectively, and the integral is over the volume of a single voxel. Eq. (5) can be expressed in terms of displacement components according to the following equation:

$$E_n = \frac{1}{2} u_{rp}^T D_{rp,sq} u_{sq} \tag{6}$$

where $D_{rp,sq}$ represents the stiffness matrix and $u_{rp}$ represents the $p$'th component of the displacement at the $r$'th node. Eqs. (4) indicates that the strain tensor of any voxel can be approximately represented as the sum of the mechanical and apparent strains, which is applied to the displacement vector at each node as $u_{rp} = U_{rp} + \delta_{rp}^{boundary} + \delta_{rp}^{apparent}$. Here, $U_{rp}$ represents the displacement vector set by the neighboring voxels, $\delta_{rp}^{boundary}$ is a correction vector set by the boundary, and $\delta_{rp}^{apparent}$ is the correction vector owing to the dissolution-induced apparent strain in the microstructure. Adopting the work of (Li et al., 2017; Li et al., 2016), if we combine the two correction vectors as $\delta_{rp} = \delta_{rp}^{boundary} + \delta_{rp}^{apparent}$, and insert it into Eq. (6), we obtain for a given voxel (Garboczi, 1998; Garboczi et al., 1999)

$$E_n = \frac{1}{2} u_{rp}^T D_{rp,sq} u_{sq} + b_{rp} u_{rp} + C$$

where

$$b_{sq} = \delta_{rp} D_{rp,sq} \text{ and } C = \frac{1}{2} \delta_{rp} D_{rp,sq} \delta_{sq}$$

(7)

The exact displacement solution to Eq. (7) can be obtained using Simpson's rule of $\frac{\partial E_n}{\partial u_{rp}} = A_{rp,sq} u_{sq} + b_{rp}$, where $A_{rp,sq}$ is the Hessian matrix built up from the stiffness matrices of the voxels/elements.

Considering the infinitesimal strain in this work, the stress and strain distributions in the microstructure were obtained by minimizing the total stored mechanical energy using FEM (Garboczi, 1998). Owing to the limitations of the computational scheme, only a strain-controlled periodic boundary condition can be applied to the microstructures. In this case, Boltzmann's superposition principle was used to predict the temporal evolution of the creep strain of the microstructure due to the constant external stress (Christensen, 2012; Wineman & Rajagopal, 2000). First, the composite was exposed to constant periodic volumetric strain to obtain the average hydrostatic stress in the composite. Then this periodic volumetric strain is applied to the composite at different loading ages to calculate the spatially averaged hydrostatic stress as a function of time. As a result of the dissolution and precipitation processes, stress can relax (because of dissolution) or increase (because of precipitation). If both dissolution and precipitation occur, the dominant process determines the stress relaxation or stress increase. If dissolution is the dominant process, additional strain must be applied at each time step to maintain the stress constant. Boltzmann's superposition principle was used to achieve this additional strain according to the following equation (Li et al., 2016)

$$if\ i = 1, \varepsilon(t_i) = \frac{-\sigma_0}{3K_{t_i}(t_i)}$$

$$if\ i \neq 1, \varepsilon(t_i) = \frac{\sigma_0 - \sum_{k=1}^{i-1}[\varepsilon(t_k)K_{t_k}(t_i)]}{3K_{t_i}(t_i)} \quad (8)$$

where $\sigma_0$ represents the maintained constant averaged hydrostatic stress, $i$ is the number of time steps, and $i = 1$ corresponds to the initial time step when the external load is applied. The body elastic deformation occurs immediately, $\varepsilon(t_i)$ is the additional linear strain required to attain the constant stress at the boundary, and $K_{t_i}(t_i)$ is the apparent VE/VP bulk modulus of the composite at time $t_i$ when loaded at age $t_k$. The total strain, which is the sum of the elastic and creep strains as a function of time, can then be obtained by $\varepsilon_{total}(t_i) = \sum_{k=1}^{i} \varepsilon(t_k)$, where $\varepsilon_{total}(t_i)$ the total resultant strain of the composite at any time $t_i$.

## 3. RESULTS AND DISCUSSION

For both $CO_2$ and $N_2$ samples, at the first time step ($t_i = 0$), a hydrostatic strain of 130 µm was applied such that the averaged hydrostatic stress was the same as the experimental pressure condition, that is, 12.5 MPa. Multiple iterations, starting from a very small strain value, were performed to determine the correct strain measurement corresponding to a pressure of 12.5 MPa. Fig. 9 shows the creep strain contours and the normalized average creep strain versus the distance from the reacted surface for the $CO_2$ and $N_2$ samples after 14 days of reaction. The average creep strain values were normalized using the creep strain in the unreacted region (Fig. 9c). These results signify the substantial role of the dissolution/precipitation processes on the viscous behavior of rocks subjected to reactive environments.

The spatial distribution of creep deformations under both $N_2$ and $CO_2$ conditions is non-uniform and depends on the extent and spatial variability of the reaction. The results showed that the rock that reacted under the $N_2$ condition experienced a very high creep strain limited to a narrow zone of approximately 200 µm close to the reacted surface. This was due to the softer response of the material as a result of weathering and washing of the grains close to the reacted surface, as well as the precipitation of weak Al-rich phases within that zone. In contrast, for the $CO_2$ condition, the time-dependent deformation spans an area of approximately 800 µm, corresponding to the severity of the mechanical alteration of different material phases owing to chemical reactions.

Under $CO_2$ condition, a low creep strain was observed close to the reacted surface until a depth of 100 µm, owing to the precipitation process. The highest amount of creep strain in the $CO_2$ sample occurred 500–600 µm away from the reacted surface, resulting from the low modulus of clay and quartz at that location. This can be explained by the fact that at the scale of nanoindentation performed in this study, the probed microvolume is composed of a mineral phase intermixed with porosity and organic matter (Abedi, Slim, Hofmann, et al., 2016; Abedi, Slim, & Ulm, 2016; Prakash et al., 2021). Therefore, the results obtained by indentation were interdependent on the small porosity between the mineral grains. As discussed in Sec. 2.2 based on the results shown in Fig. 2a, both the clay-rich and quart-rich phases weakened and became more porous compared to the unreacted phases owing to the dissolution of the mineral grains. However, part of the porosity in these phases is filled by precipitation closer to the reacted surface, resulting in increased modulus of these phases in the vicinity of the exposed surface.

Based on the concept of pore-size-controlled solubility (PCS), the level of supersaturation necessary for crystal growth increases with the confinement pressure (Emmanuel & Ague, 2009; Varzina et al., 2020) . This phenomenon, which is similar to the Ostwald ripening (Nabika et al., 2019; Voorhees, 1985) process, in which large crystals grow at the cost of smaller crystals with higher solubility, limits the precipitation of secondary minerals in the nano-porosity between mineral grains. Based on the results presented in Fig. 9, the effect of precipitation on the small pores between grains diminishes at a distance of 500–600 m away from the reacted surface, which causes a lower modulus of clay and quartz, leading to the elevated creep deformation observed at this point. Therefore, the interplay between dissolution and precipitation dictates the amount of time-dependent deformation of rocks, especially when exposed to harsh environments, such as acidic environment caused by $CO_2$ and brine under high-pressure and high-temperature conditions. This also indicates that the evolution of the pore-size distribution induced by chemo-mechanical interactions plays a key role in the creep properties of rock materials.

To understand the contribution of individual minerals to the total creep strain, the average creep strain of clay-, quartz-, and feldspar-rich minerals was plotted against the distance from the exposed surface for the $CO_2$ condition after 14 days of reaction (Fig. 10). Clay-rich phases exhibited the highest creep strain, followed by quartz-rich phases, and the lowest creep strain was observed in feldspar-rich phases. Because of the dissolution of feldspar and precipitation of secondary minerals, clay has a very high volume fraction adjacent to the reacted surface; it has lower stiffness compared to quartz, and its modulus has diminished the most in the reacted region. The combination of these effects results in a higher contribution of clay minerals to creep deformation. This plot highlights the effect of the composition of the shale sample on creep strain. A very high volume fraction of clay minerals can increase creep deformation and have more viscous effects on the rock.

The analysis was further extended to obtain the creep strain under the $CO_2$ condition when the reaction depth grew over time. This part of the analysis does not account for the mechanical degradation/alteration of the material phases over time and only explores the role of the growth of the reaction depth on creep deformation. Therefore, the results may underestimate creep deformation for a longer reaction duration. Further experimental studies are planned to be conducted for longer durations to capture the effects of both microstructural and mechanical changes on creep deformation as the reaction progresses.

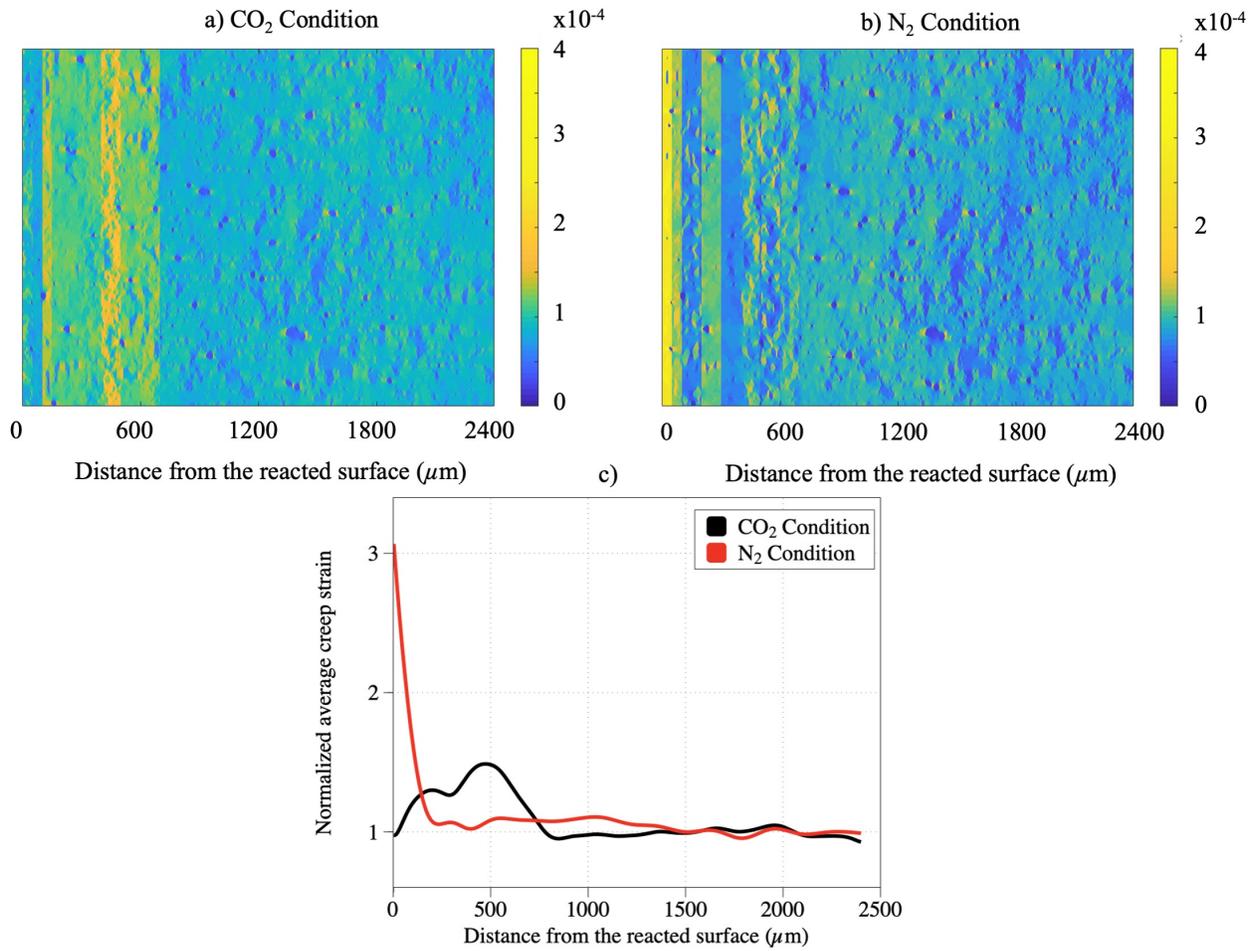

Figure 9- Contours of creep strain field for a) $CO_2$ condition, and b) $N_2$ condition after 14 days of reaction. c) Normalized average creep strain vs distance from the reacted surface. The reacted surface is on the left side of the images.

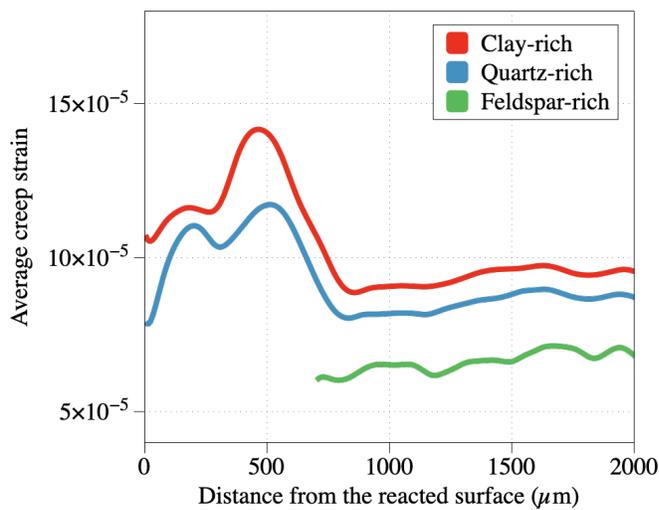

Figure 10- Average creep strain of clay-, quartz- and feldspar-rich phases vs distance from the exposed surface for the sample reacted under $CO_2$ condition for 14 days.

As mentioned in Sec. 2.3.1, for this analysis, a microstructural model was developed with the assumption that the reaction depth is a function of time, and it follows Fick's law (Eq. 1). For the subsequently generated microstructures, the same boundary conditions were applied to the updated microstructure. Stress relaxation was observed at each time step, indicating the dominance of the dissolution reaction. The creep strain at all time steps was calculated using Boltzmann's equation (Eq. 8). Fig. 11 shows the predicted creep of Permian shale, excluding the instantaneous elastic deformation due to loading. Each creep strain value in the corresponding figure represents the total creep at the end of the loading period. As depicted in Fig. 11, the shale sample continues to deform under constant loading due to the progression of the reaction depth, even without considering further mechanical alteration of the material phases as the reaction evolves. Fig. 12 shows the decrease in the VE/VP bulk modulus of Permian shale with increasing reaction depth.

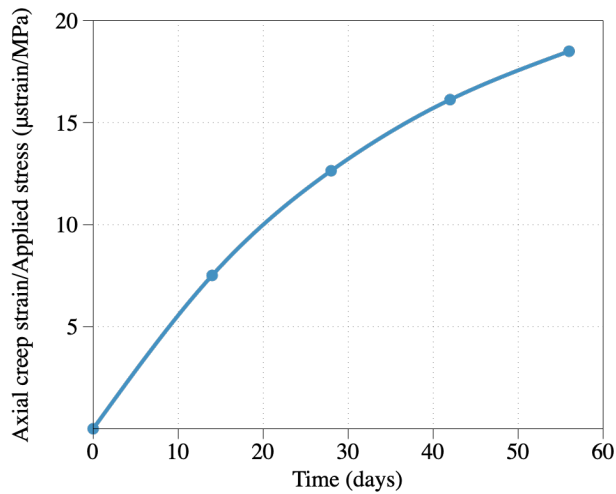

Figure 11- Axial creep strain of Permian sample with increase in reaction depth over time for the sample reacted under $CO_2$ condition.

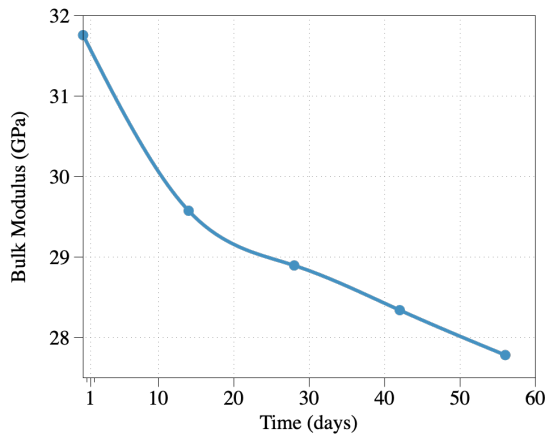

Figure 12- Viscoelastic bulk modulus of Permian shale with increase in reaction depth over time for the sample reacted under $CO_2$ condition.

Batch-type experiments performed in the laboratory have limitations on the pressure inside the reaction chamber; hence, in this study, the experiments were performed at a pressure of 1800 psi. However, shale rocks can experience very high pressure values deep inside the rock bed. An FEM analysis was performed to compute the creep strain at higher pressure values. Simulations were performed at 24.8, 49.6, and 99.28 MPa (3600, 7200, and 14400 Psi). Although both thermodynamics and reaction kinetics are expected to be affected by the state of pressure applied to rocks, for simplification in the analysis, the reactivity and alteration in the mechanical properties of the shale minerals were assumed to be constant with an increase in pressure. At all these pressure values, the creep strain was calculated for all loading ages (14, 28, 42, and 56 days), as it was calculated for the 12.5 MPa condition. The results showed that the creep strain increased with increasing applied pressure (Fig. 13). Further experimental and modeling work is required to quantify the real extent of the reaction over time as a function of pressure to obtain a more realistic picture of the creep strain as the reaction progresses with time.

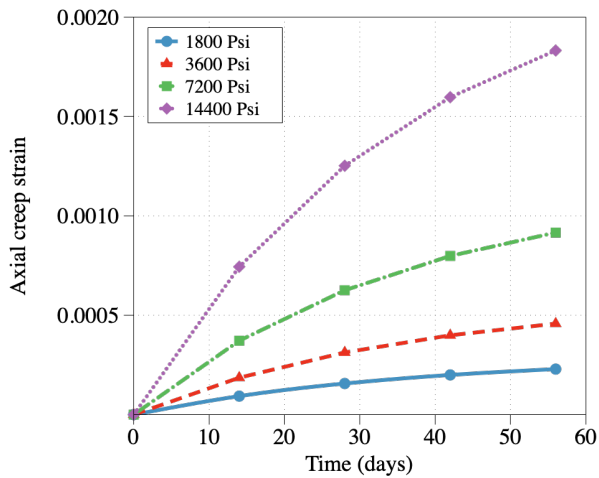

Figure 13- Change in axial creep strain with increase in the reaction depth over time at different loading for the sample reacted under $CO_2$ condition.

Notably, the bulk modulus obtained reflects the viscoelastic response of an area with the size of 2400 μm × 1800 μm of a shale rock subjected to hydrostatic loading. This viscoelastic response is dependent on the size of the simulation box. Here, we focused on the region in the vicinity of the reacted surface.

## 4. CONCLUSION

A computational scheme integrated with experimentally measured microstructural and mechanical data was adopted to calculate the time-dependent creep strain of shale rock induced by chemo-mechanical loading. A time-dependent microstructural model and time-stepping finite element model were coupled to build the computational framework.

The microstructural model made use of results from micro-CT imaging and coupled nanoidnetation/EDS techniques to obtain the evolution of the microstructure of the samples owing to chemo-mechanical loading. The approach was applied to Permian shale samples to forecast viscoelastic behavior when subjected to $CO_2$- and $N_2$-rich brine under high-pressure and high-temperature conditions. The results indicated that when it comes to the effect of rock-fluid interaction on the time-dependent mechanical properties of rocks, the effect of dissolution/precipitation processes during the reaction is substantial toward the viscous properties of rocks.

For both the $CO_2$ and $N_2$ conditions, the spatial distribution of the creep deformations was not uniform and depended on the extent and spatial variability of the reaction. The sample that reacted under $N_2$ condition showed a very high creep strain limited to a narrow zone (i.e., 200 µm) close to the reacted surface. This was due to weathering and washing of the grains close to the reacted surface, as well as the precipitation of weak Al-rich phases within that zone.

In contrast, for the sample reaction under $CO_2$ conditions, the time-dependent deformation caused by the dissolution/precipitation processes extended to an area of approximately 800 µm, which corresponds to the severity of the mechanical alteration of different material phases. For this sample, reduced creep deformation was observed near the reacted surface to a depth of 100 µm owing to the precipitation process. The creep deformation reached a maximum 500–600 µm away from the reacted surface, where the effect of precipitation on the small pores between mineral grains diminished. This observation highlights the determining role of the interplay between the dissolution and precipitation processes and the resulting pore size distribution on the VE/VP behavior of rocks when exposed to reactive environments.

The creep strain rate was obtained for different mineral phase components in different reaction zones. The clay-rich phases exhibited the highest creep strain owing to the very high volume fraction of clay close to the reacted surface and its lower stiffness compared to quartz. Further extension of the analysis to consider the growth of the reaction depth over time for $CO_2$ exposed sample showed a continuous reduction in the VE/VP bulk modulus of the microstructure with time.

**Acknowledgment**

Acknowledgment is made to the National Science Foundation (Grant CMMI-2045242) and to the donors of the American Chemical Society Petroleum Research Fund (PRF 60545- ND9) for supporting this work.

**Ethics Declaration –** Conflict of Interest

The authors declare that there is no conflict of interest associated with this publication.